\documentclass[prd,twocolumn,amsmath,amssymb,floatfix,nofootinbib]{revtex4}
\usepackage{graphicx,dcolumn,booktabs}
\usepackage{amsmath}
\usepackage{amsfonts}
\usepackage{amssymb}
\usepackage{times}
\usepackage{amsmath, amssymb}

\begin{document}
\title{Re-study on the contribution of scalar potential and
 spectra of $c\bar c$, $b\bar b$ and $b\bar c(\bar b c)$ families}

\author{ Xu-Hao Yuan $^{1}$   \footnote{segoat@mail.nankai.edu.cn},
         Hong-Wei Ke $^{2}$  \footnote{khw020056@hotmail.com, corresponding author}
         Yi-Bing Ding $^3$
         Xue-Qian Li $^{1}$  \footnote{lixq@nankai.edu.cn} }

\affiliation{
  $^{1}$ School of Physics, Nankai University, Tianjin 300071,
  China\\
  $^{2}$ School of Science, Tianjin University, Tianjin 300072,
  China\\
  $^3$ College of Physics Sciences, Graduate University,
   Chinese Academy of Sciences, Beijing 100049 }

\begin{abstract}
\noindent We indicated in our previous work that for QED the role
of the scalar potential which appears at the loop level is much
smaller than that of the vector potential and in fact negligible.
But the situation is different for QCD, one reason is that the
loop effects are more significant because $\alpha_s$ is much
larger than $\alpha$, and secondly the non-perturbative QCD
effects may induce a sizable scalar potential. In this work, we
phenomenologically study the contribution of the scalar potential
to the spectra of charmonia, bottomonia and $b\bar c(\bar b c)$
family. Taking into account both vector and scalar potentials, by
fitting the well measured charmonia and bottomonia spectra, we
re-fix the relevant parameters and test them by calculating other
states of not only the charmonia, bottomonia, but also further the
$b\bar c$ family. We also consider the Lamb shift of the spectra.
\end{abstract}

\maketitle

\section{Introduction}

The potential model has been proposed to evaluate the spectra of
the quantum systems composing of heavy flavors, such as charmonia
and bottomonia for many years \cite{Quigg:1979vr}. The subject on
heavy quarkonia was thoroughly discussed in an enlightening paper
\cite{Brambilla:2004wf}. Even though one calculates the binding
energy in terms of the non-relativistic Schr\"odinger equation, it
is indeed a reasonable framework for such heavy resonant states.
The main physics, no matter it is induced by the standard model
(SM) or new physics beyond the SM, is included in the potential.
In phenomenology, the potential contains two pieces, the Coulomb
potential which is induced by the gluon exchange and the
confinement which may come from the non-perturbative QCD effects
or other sources. It was suggested that due to the symmetry
consideration \cite{chen:2008}, for the QED case, if the Dirac
equation has a higher symmetry degree than the corresponding
Schr\"odinger equation, the Coulomb potential should have both
scalar and vector parts with equal fractions. However, for such a
combination the hydrogen potential would not possess the coupling
of orbital angular momentum and spin (${\bf L}\cdot{\bf S}$ ) and
it definitely contradicts to the reality.

In our previous work \cite{Ke:2009mx}, we analyzed the source of
the scalar and vector potentials for the QED, namely the vector
potential is induced by one-photon exchange for the vector-like
gauge theory QED, while the scalar part must be coming from the
loop effects. Therefore, for the QED case, the fraction of the
scalar potential is very suppressed because the coupling
$\alpha\sim 1/137$ is small and does not have substantial
contribution to the spectrum of hydrogen. This conclusion is
consistent with the precise measurement on the spectra of
hydrogen-like atoms. However, for the QCD case, the situation is
very different. First because the strong coupling $\alpha_s$ is
much larger than $\alpha$, the loop suppression is not as much as
for the QED and sometimes the NLO effect of QCD can even exceed
the leading order\cite{Zhang:2006ay,Gong:2009kp}. Secondly the
non-perturbative effects may also change the whole scenario. Thus
for QCD, the scalar potential may play an important role and its
contribution may especially manifest at values of the spectrum
line splitting due to the coupling of orbital angular momentum and
spin. That was also suggested by some
authors\cite{Leviatan:2003wy}.

Because the non-perturbative QCD effects cannot be reliably
derived from any underlying theory so far, we prefer to introduce
phenomenological parameters to manifest their roles and the
parameters are fixed by fitting the well-measured data. Indeed the
parameters not only contain the contribution of the
non-perturbative effects, but also that from the higher orders of
the perturbative effects. In our last work \cite{Yuan:2010jy}, we
introduced two more parameters to account for the scalar piece in
the potential and then re-fitted the spectra of charmonia. The
fitting is indeed improved comparing with that including only the
vector piece. Thus for further investigating the involvement of
scalar potential, we use the same strategy adopted in
\cite{Yuan:2010jy} to evaluate the spectra of the bottomonia
($b\bar b$) and then the various resonant states of the $b\bar
c(\bar b c)$ system. Namely, we write $U(r)=V(r)+S(r)$ and the
vector potential is $V(r)=-c~C_F\alpha_s/r+d\kappa^2r$, while the
scalar one is $S(r)=-(a-c)C_F\alpha_s/r+(b-d)\kappa^2r$ with
$a,b,c,d$ are four parameters to be determined. The induced terms,
such as $\bf L\cdot\bf S$ coupling, ${\bf S}_1\cdot {\bf S}_2$
coupling and the spin-independent corrections etc. are given in
the very enlightening work by Lucha et al.
\cite{Lucha:1991it,Lucha:1991vn,Ding:1991vu}. Substituting all the
expressions into the Schr\"odinger equation, we may obtain the
spectra of heavy quark-antiquark systems.

Besides, the QED theory predicts the Lamb shift which is due to
the vacuum effects. In QM, it only shifts the $\mathrm S$-wave
spectra because in the non-relativistic limit, it is proportional
to $\delta({\bf r})$, but by the quantum field theory, the ${\bf
L}\neq0$ states are also affected. In other words, by considering
the Lamb shift, the positions of the spectra would deviate from
that obtained without the Lamb shift. For the hadron case, the
governing theory is QCD which also induces the Lamb shift
\cite{Hoang:2001rr}, and in this work, we include its
contribution. It is noticed that, as the phenomenological
parameters which are determined by fitting data are introduced,
all higher order effects should also be automatically involved, so
it seems that there is no need to consider the Lamb shifts which
are induced by the loop effects. In fact, it is. However, as
calculating the form factors of the hadronic transition matrix
elements or obtaining the parton distribution functions, we always
wish to squeeze the uncontrollable parts which are not calculable,
such as the non-perturbative contributions, as small as possible.
Similarly, here we include the NLO or even NNLO corrections i.e.
the Lamb shifts and re-fit the parameters which are indeed not
derivable.

Considering the Lamb shift, there is a byproduct, which may be
very enlightening for understanding the theory. Obviously, only
the products $c\alpha_s$ and $d\kappa$ in the potential matter,
but not $c$, $d$, $\alpha_s$ and $\kappa$ separately. However, for
explicitly showing the roles of the scalar and vector pieces, we
adopt a special strategy.

When the Lamb shift is taken into account, the situation may
change slightly. In the expression of the Lamb shift, there is an
ultraviolet divergent term which includes the renormalization
scale $\mu$. Meanwhile the running coupling $\alpha_s$ also
depends on the scale. The authors of \cite{Titard:1993nn} suggest
an effective method to deal with the divergence and meanwhile fix
the value of $\alpha_s$ (see the text, where we introduce the
method in some details for the readers' convenience). It is noted
that $\mu$ is a complicated function of $\alpha_s$, quark mass
($m_b$ or $m_c$), and the principal quantum number $\mathrm n$.
Taking a special way to determine $\mu$ which is corresponding to
adopting a special renormalization scheme, and considering the
dependence of $\alpha_s$ on $\mu$, one can eventually find the
value of $\alpha_s$ for a certain flavor ($b$ or $c$) and a given
principal quantum number. For example, for $\Upsilon(1\mathrm S)$,
we have $\alpha_s=0.218$ for $m_b=4.8$ GeV and the scale-parameter
$\Lambda=0.2$ GeV. Amazingly, this value is quite close to that
adopted in literature by fitting the $\Upsilon$ spectra.

By contrast, the confinement term $\kappa r$ is fully coming from
the non-perturbative QCD effects and cannot be theoretically
derived so far. Thus we adopt the value given in \cite{Ke:2009sk}.

With all the inputs, we calculate the spectra of the $c\bar c$,
$b\bar b$ and $\bar b c(b\bar c)$ systems.

This paper is organized as follows. In Sec.\ \ref{sec-2} and
\ref{sec-3}, we introduce the generalized Breit-Fermi Hamiltonian
and the Schr\"odinger equation for the $b\bar{b}$ bound states:
$\Upsilon(1\mathrm S)$, $\chi_{b0}(1\mathrm P)$,
$\chi_{b1}(1\mathrm P)$, $\chi_{b2}(1\mathrm P)$ and
$\Upsilon(2\mathrm S)$. Then we numerically solve the
eigen-equations for these bound states and fix the parameters as
we did for dealing with the charmonia family in our previous work.
In Sec.\ \ref{sec-4}, the Lamb shift is taken into account and
another set of the parameters is given to improve our predictions.
In Sec.\ \ref{sec-5}, we give the spectra of the $\bar b c(b\bar
c)$ mesons. The last section is devoted to our conclusion and
discussion.

\section{The Generalized Breit-Fermi Hamiltonian and Sch\"ordinger
 equation}\label{sec-2}
The generalized Breit-Fermi Hamiltonian was given in Refs.
\cite{Lucha:1991it,Lucha:1991vn,Ding:1991vu} as
\begin{subequations}\label{B-F Hamilton}
\begin{eqnarray}
 &&H=H_0+H_1+\dots\ ,\\
 &&H_0={p^2\over m}+2m+S(r)+V(r)\ ,
\end{eqnarray}
\begin{eqnarray}
 H_1&=&H_\mathrm{sd}+H_\mathrm{si}\ ,\\
 H_\mathrm{sd}&=&H_\mathrm{ls}+H_\mathrm{ss}+H_\mathrm{t}\nonumber\\
 &=&{1\over2m^2r}\left(3V'-S'\right)
 {\bf L}\cdot({\bf S}_1+{\bf S}_2)
  +{2\over3m^2}{\bf S}_1\cdot{\bf S}_2\nabla^2V(r)\nonumber\\
 & &+{1\over12m^2}\left({1\over
  r}V'-V''\right){\bf S}_{12}\ ,\\
 H_\mathrm{si}&=&-{p^4\over4m^3}
  +{1\over4m^2}\left\{{2\over r}V'(r)\cdot {\bf L}^2
   +[p^2,V-r V']\right.\nonumber\\
 & &+\left.2(V-r V')p^2+{1\over2}\left[{8\over
  r}V'(r)+V''-r V'''\right]\right\}
\end{eqnarray}
\end{subequations}
where, $V$ and $S$ stand as the vector and scalar potentials and
$H_\mathrm{si}$ and $H_\mathrm{sd}$ represent the spin-independent
and spin-dependent pieces respectively. For the linear confinement
piece we adopt the Cornell potential\cite{Eichten:cornell}. Thus
the total potential at the lowest order reads
\begin{subequations}\label{v-s}
\begin{eqnarray}
 U(r)=V(r)+S(r)=-aC_F{\alpha_s\over r}+b\kappa^2r
\end{eqnarray}
\mbox{where},
\begin{eqnarray}
 \left\{
  \begin{array}{l}
   V(r)=-c~C_F\alpha_s/r+d\kappa^2r\\
   S(r)=-(a-c)C_F\alpha_s/r+(b-d)\kappa^2r
  \end{array}
 \right.
\end{eqnarray}
\end{subequations}
With the hamiltonian (\ref{B-F Hamilton}) and the potential
(\ref{v-s}), one can solve the Schr\"odinger equation
\begin{eqnarray}\label{sch.}
 H\Psi(r)=(H_0+H_1)\Psi(r)=(E+2m)\Psi(r).
\end{eqnarray}

We can transform the radial wave function into $R(x)$ with a
dimensionless variable: $x=\kappa r$, then the reduced radial
equation is written as\footnote{The standard form of the radial
equation can be easily found in Ref.\cite{Cai:2003}, and the
method to make it dimensionless is borrowed from
Ref.\cite{Silbar:2010}.}
\begin{subequations}\label{radl.eqn}
\begin{eqnarray}
 {d^2\over dx^2}u(x)=A(x)u(x),
\end{eqnarray}
where,
\begin{eqnarray}
 A(x)&=&-\tilde{m}\left(\tilde{E}-\tilde{U}(x)-\tilde{H}_1\right)
  +{\mathrm l(\mathrm l+1)\over x^2}\nonumber\\
 & &-{1\over4}\left(\tilde{E}-\tilde{U}(x)\right)^2
\end{eqnarray}
with
\begin{eqnarray}
 \left\{
  \begin{array}{l}
   \tilde{m}=m/\kappa,\quad \tilde{E}=E/\kappa,\\
   \tilde{H}_1=H_1/\kappa,~ \tilde{U}(x)=U(x)/\kappa.\\
  \end{array}
 \right.
\end{eqnarray}
In the simplified potential form (\ref{sch.}), the approximation
\begin{eqnarray}
 p^4\sim\Big[m\left(E-U(r)\right)\Big]^2
\end{eqnarray}
\end{subequations}
is used. For the legitimacy of applying this approximation in the
calculation and the error degree brought up in the numerical
values are briefly discussed in the appendix.

\section{The Energy Gap Function for The $b\bar{b}$ bottomonia and The numerical
Results without taking into account the Lamb shift}\label{sec-3}

The radial equation (\ref{radl.eqn}) can be solved in terms of the
method so-called ``The iterative numerical process'' which is
introduced in literatures (for example, see
\cite{Silbar:2010,Cai:2003}). We have improved this method, and
then fix the parameters $a$, $b$, $c$, $d$ by fitting the well
measured spectra of bottomonia. In our previous work
\cite{Yuan:2010jy}, we explain the reason for the choice of the
input for charmonia. However, for bottomonia, the situation is
slightly different and the masses of $\Upsilon(1\mathrm S)$,
$\chi_{b0}(1\mathrm P)$, $\chi_{b1}(1\mathrm P)$,
$\chi_{b2}(1\mathrm P)$ and $\Upsilon(2\mathrm S)$ are chosen for
the obtaining the values of $a$, $b$, $c$ and $d$. Similar to the
procedure we took in our previous work \cite{Yuan:2010jy}, instead
of directly fitting the masses, we construct a series of relations
which should be fitted:\\
\begin{eqnarray}\label{e-g}
 \left\{
  \begin{array}{l}
   m\left[\Upsilon(2\mathrm S)\right]
    -m\left[\Upsilon(1\mathrm S)\right]
     =E\left[2^3\mathrm S_1\right]
      -E\left[1^3\mathrm S_1\right];\\
   m\left[\Upsilon(2\mathrm S)\right]
    -m\left[\chi_{b0}(1\mathrm P)\right]
     =E\left[2^3\mathrm S_1\right]
      -E\left[1^3\mathrm P_0\right];\\
   m\left[\Upsilon(2\mathrm S)\right]
    -m\left[\chi_{b1}(1\mathrm P)\right]
     =E\left[2^3\mathrm S_1\right]
      -E\left[1^3\mathrm P_1\right];\\
   m\left[\Upsilon(2\mathrm S)\right]
    -m\left[\chi_{b2}(1\mathrm P)\right]
     =E\left[2^3\mathrm S_1\right]
      -E\left[1^3\mathrm P_2\right]\ ,
  \end{array}
 \right.
\end{eqnarray}
where, $E\left[\mathrm n_\mathrm r^{2\mathrm s+1}\mathrm l_\mathrm
j\right]$ represents the eigen-values of the radial equations
(\ref{radl.eqn}) with various quantum numbers $\mathrm n_\mathrm
r$, $\mathrm j$, $\mathrm l$, and $\mathrm s$ for the bottomonia.
Because the parameters $a$, $b$, $c$ and $d$ are involved in the
potential (\ref{v-s}), $E\left[\mathrm n_\mathrm r^{2\mathrm
s+1}\mathrm l_\mathrm j\right]$ must be functions of these
parameters. $m[\mbox{meson}]$ are the masses of the individual
states which are shown in the following Tab.\
\ref{tab1}\cite{Amsler:2008zzb}.

Sequentially, the parameters $a$, $b$, $c$ and $d$ are obtained by
solving Eqs.(\ref{e-g}). By employing the Newton's iterative
method (The details about the numerical method can be found in
Ref. \cite{Press:2007}.), we have achieved:
\begin{eqnarray}\label{abcd}
 a=1.2165,~b=1.2988,~c=0.8686,~d=0.5886
\end{eqnarray}

Here we set $\alpha_s=0.284$ and $\kappa=0.42$ GeV which seem
somehow different from the values given in
literature\cite{Ke:2009sk,Ding:1988pg,Bali:1992}. But as noticed,
the deviation may be included in the phenomenological parameters
$a$, $b$, $c$ and $d$. The choice of $\alpha_s$ has another reason
which is associated with our treatment of the contribution of the
Lamb shift (see next section), anyhow this value is not very far
apart from that given in literature.

Given  $a$, $b$, $c$ and $d$ in (\ref{abcd}), the masses of the
bottomonia states are determined as:

\begin{eqnarray}\label{f-r}
 M(\mathrm n_\mathrm r^{2\mathrm s+1}\mathrm l_\mathrm j)
  =E\left[\mathrm n_\mathrm r^{2\mathrm s+1}\mathrm l_\mathrm j\right]
   +E_0
\end{eqnarray}
where, $E_0$ is the zero-point energy:
\begin{eqnarray}\label{zpe}
 \left.
  \begin{array}{l}
   E_0=m[\Upsilon(1\mathrm S)]-E[1^3\mathrm S_1]
  \end{array}
 \right.
\end{eqnarray}
and the final results are shown in the Tab.\ \ref{tab1} below.

\begin{center}
\begin{table}[!h]
\caption{The mass spectra for the bottomonia states
 (in GeV), with $m_b=4.8\mathrm {GeV}$.
 The $M_\mathrm{EXP}$ is the value of the mass given in
  PDG\cite{Amsler:2008zzb}.} \label{tab1}
\begin{tabular}{lcclcc}
 \toprule[1pt]
  meson &
   $M_\mathrm{EXP}$ &
    Prediction &
     meson & $M_\mathrm{EXP}$ & Prediction
    \\ \midrule[0.5pt]
    $\eta_b(1^1\mathrm{S}_0)$ & 9.3020 & 9.4124 &
     $\eta_b(2^1\mathrm{S}_0)$ &   & 9.9932
    \\
    $\Upsilon(1^3\mathrm{S}_1)^\mathrm{fit}$ & 9.4603 & 9.4603 &
     $\Upsilon(2^3\mathrm{S}_1)^\mathrm{fit}$ & 10.0233 & 10.0233
    \\
    $\chi_{b0}(1^3\mathrm{P}_0)^\mathrm{fit}$ & 9.8594 & 9.8594 &
     $\Upsilon(1^3\mathrm D_1)$ & 10.1611 & 10.1607
    \\
    $\chi_{b1}(1^3\mathrm{P}_1)^\mathrm{fit}$ & 9.8928 & 9.8928 &
     $\Upsilon(1^3\mathrm D_2)$ &   & 10.1719
    \\
    $h_b(1^1\mathrm{P}_1)$ &   & 9.8897 &
     $\Upsilon(1^3\mathrm D_3)$ &   & 10.1827
    \\
    $\chi_{b2}(1^3\mathrm{P}_2)^\mathrm{fit}$ & 9.9122 & 9.9122 &
     $\Upsilon(3^3\mathrm{S}_1)$ & 10.3552 & 10.3949
    \\
 \bottomrule[1pt]
\end{tabular}
\end{table}
\end{center}

Explicitly, in the process, the masses of the mesons with
superscript ``fit'' are taken as inputs to obtain the parameters
and then the masses of other states in the table are predicted.

For readers' convenience and a clear comparison, we also list the
results for the charmonia which were obtained in our previous work
\cite{Yuan:2010jy}.

\begin{center}
\begin{table}[!h]
\caption{The mass spectra for the charmonia states
 (in GeV), with $m_\mathrm{c}=1.84\mbox{GeV}$.
 The $M_\mathrm{EXP}$ is the value of the mass given in
  PDG\cite{Amsler:2008zzb}.}\label{tab-cc}
\begin{tabular}{lcclcc}
 \toprule[1pt]
  meson &
   $M_\mathrm{EXP}$ &
    Prediction &
     meson & $M_\mathrm{EXP}$ & Prediction
    \\ \midrule[0.5pt]
    $\eta_c(1^1\mathrm{S}_0)$ & 2.9803 & 3.0189 &
     $\chi_{c2}(1^3\mathrm{P}_2)^\mathrm{fit}$ & 3.5562 & 3.5564
    \\
    $J/\psi(1^3\mathrm{S}_1)^\mathrm{fit}$ & 3.0969 & 3.0969 &
     $\eta_c(2^1\mathrm{S}_0)^\mathrm{fit}$ & 3.6370 & 3.6370
    \\
    $\chi_{c0}(1^3\mathrm{P}_0)^\mathrm{fit}$ & 3.4148 & 3.4148 &
     $\psi(2^3\mathrm{S}_1)$ & 3.6861 & 3.6861
    \\
    $\chi_{c1}(1^3\mathrm{P}_1)^\mathrm{fit}$ & 3.5107 & 3.5107 &
     $\psi(3^3\mathrm{S}_1)$ &   & 4.1164
    \\
    $h_c(1^1\mathrm{P}_1)$ & 3.5259 & 3.5100 & & &\\
     \bottomrule[1pt]
\end{tabular}
\end{table}
\end{center}
with
\begin{eqnarray}\label{abcd-c}
 a=1.1715,~b=1.2250,~c=0.8087,~d=0.5291
\end{eqnarray}

\section{The mass spectra of the bottomonia as the Lamb
shift is taken into account}\label{sec-4}

As well known, the Lamb shift is due to the vacuum fluctuation and
may cause sizable effects on the meson spectra. Indeed, the QED
Lamb shift may not be very significant because of smallness of the
fine structure constant $\alpha$ \cite{Greiner:1992bv}, in fact,
it just reaches order of $10^{-7}$ eV, but for the QCD case, the
situation is different.

In the previous section the effects of the Lamb shift are not
included in the eigen-energy (\ref{radl.eqn}). Thus in this
section, we take the Lamb shift into account. However, we do not
introduce the Hamiltonian induced by the Lamb shift into the
differential equation because the corresponding pieces are very
complicated. Instead, according to the traditional method, we
calculate the effects in terms of the wavefunctions obtained with
the original Hamiltonian, i.e. accounting
$<\Psi|H_{\mathrm{Lamb}}|\Psi>$, where $H_{\mathrm{Lamb}}$ is
obtained via the loop diagrams and simply add the estimated values
into the binding energies of various states. Including the Lamb
shift effects in the expressions of the spectra, we re-fit the
data to obtain the parameters $a$, $b$, $c$ and $d$ again and
predict the mass spectra of the rest resonances.

Namely, we set the masses of a bound states to be the measured
values:
\begin{equation}
2m_b+E+\Delta E_\mathrm{LM}=M_{\mathrm{EXP}},
\end{equation}
where $E$ is the solution of the eigen-equation, $\Delta
E_\mathrm{LM}$ is the energy caused by the Lamb shift and
$M_\mathrm{EXP}$ has been defined in Tab.\ \ref{tab1} and
\ref{tab-cc} already. Solving the equation, one can obtain the
parameters again.

The authors of
Ref.\cite{Hoang:2001rr,Titard:1993nn,Brambilla:1999xf} gave the
theoretical expressions for the binding energies  which involve
contributions of the Lamb shift. It is well known that the induced
Hamiltonian contributing to the Lamb shift is due to the vacuum
fluctuation and can be obtained by calculating the loop diagrams
order by order. Thus the Lamb Shift starts at
$O(\alpha_s^2)$\cite{Titard:1993nn}\footnote{We indicate that the
Lamb shift effects start from $O(\alpha_s^2)$, and it seems that
this allegation conflicts with Eq.(\ref{ls}). As noted, in the
expression Eq.(\ref{ls}) which is given in
Ref.\cite{Titard:1993nn}, $\Delta E_\mathrm{LM}$ is not the
potential derived from the loop, but the expectation value of the
potential with the wavefunctions which are solutions of the
Schr\"odinger equation containing only the Coulomb piece. Since
such solutions possess an exponential factor $\exp{-\alpha_s\mu
r}$, the expectation value of any function of $r$ as $f(r)$ should
be proportional to $\alpha_s^n\; (\mathrm n\geq 1)$, thus the
$\Delta E_\mathrm{LM}$ in Eq.(11) start from $\alpha_s^3$. But
indeed the potential pieces corresponding to the Lamb shifts
directly come from the loop diagrams which start from
$O(\alpha_s^2)$.}. The Lamb shift is:
$$\Delta E_\mathrm{LM}=<\Psi|V_{\mathrm{Lamb}}|\Psi>,$$
where $\Psi$ is the solution of the Schr\"odinger equation
containing only the Coulomb piece, can be written as :
\begin{subequations}\label{ls}
\begin{eqnarray}
 \Delta E_\mathrm{LM}[\mathrm n,\mathrm j,\mathrm l,\mathrm s]
  &=&m\Big[\Delta E(\alpha_s^3)
   +\Delta E(\alpha_s^4)+\Delta
    E(\alpha_s^5)\nonumber\\
 & &+\Delta E(\alpha_s^6)+\dots\Big]\ .
\end{eqnarray}

For illustrating the contribution of the Lamb shift to the
spectra, let us directly copy Titard's formulas
\cite{Titard:1993nn} below, where we dropped the tree-level terms
and the relativistic corrections, and we have:
\begin{eqnarray}
 \Delta E(\alpha_s^3)&=&-\alpha_s^3{C_F^2\over8\pi
 n^2}\left(2\beta_0\gamma_E+4a_1\right);\\
 \Delta
 E(\alpha_s^4)&=&-\alpha_s^4{C_F^2\over4n^2\pi^2}\left\{(a_1
 +\gamma_E{\beta_0\over2})^2
 +2\Big[\gamma_E(a_1\beta_0\right.\nonumber\\
 & &\left.+{\beta_1\over8})
 +({\pi^2\over12}
 +\gamma_E^2){\beta_0^2\over4}+b_1\Big]\right\}~,
\end{eqnarray}
\end{subequations}
where, the $\mathrm n$ in Eq.(\ref{ls}) stands for the principal
quantum number as $\mathrm n=\mathrm n_\mathrm r+\mathrm l$,
where, $\mathrm n_\mathrm r$ and $\mathrm l$ are defined in Sec.\
\ref{sec-3}. All the constants as $a_1$, $a_2$, $b_1$,
$\beta_{i}\; (i=1,2,3)$ are given in Ref.\cite{Pineda:1998} (also
see
\cite{Billoire:1979ih,Fischler:1977yf,Titard:1993nn,Peter:1996ig,Schroder:1998vy}).
Further more, Hoang et al. estimated the contribution of higher
orders up to $O(\alpha_s^5)$ and $O(\alpha_s^6)$ to the binding
energies(See \cite{Hoang:2001rr}).

When we calculate the QCD Lamb shift effects for the hadron
spectra, the potential not only contains the Coulomb piece, but
also the confinement, so that the expression would be more
complicated than that shown in Eq.(11b) and (11c). In fact, there
cannot be analytical expressions for the expectation values.

The Lamb Shift $\Delta E_\mathrm{LM}[\mathrm n,\mathrm j,\mathrm
l,\mathrm s]$ depends on the coupling constant $\alpha_s$ in
Eq.(\ref{ls}) as\cite{Titard:1993nn}:
\begin{eqnarray}
 \alpha_s(\mu^2)&=&{2\pi\over\beta_0\ln{\mu/\Lambda}}\Big\{
  1-{\beta_1\over\beta_0^2}{\ln(\ln\mu^2/\Lambda^2)\over\ln\mu^2/\Lambda^2}
  \nonumber\\
 & &+{1\over\beta_0^4\ln^2\mu^2/\Lambda^2}
  \Big[\beta_1^2\ln^2(\ln\mu^2/\Lambda^2)\nonumber\\
 & &-\beta_1^2\ln(\ln\mu^2/\Lambda^2)-\beta_1^2+\beta_2\beta_0\Big]\Big\}.
\end{eqnarray}

Using the formulas given above, one can evaluate the Lamb shift of
the charmonia states. The choice of the renormalization point
$\mu$ is suggested by Pineda et al., and ``a natural value for
this parameter '' is \cite{Titard:1993nn,Pineda:1998}:
\begin{subequations}\label{mu}
\begin{eqnarray}
 \mu={2\over \mathrm n a_B}
\end{eqnarray}
where,
\begin{eqnarray}
 a_B&=&{2\over m C_F \tilde{a}_s}\\
 \tilde{\alpha}_s(\mu^2)&=&\alpha_s\Big\{1
  +(a_1+{\gamma_E\beta_0\over2}){\alpha_s\over\pi}
   \Big[\gamma_E\left(a_1\beta_0+{\beta_1\over8}\right)
  \nonumber\\
 & &+({\pi^2\over12}+\gamma_E^2){\beta_0^2\over4}+b_1\Big]
     {\alpha_s^2\over\pi^2}\Big\}.
\end{eqnarray}
\end{subequations}

In the expression of the newly derived Hamiltonian there is a term
$\ln {2\alpha\mu r}/r$ (after a Fourier transformation from the
momentum space to the configuration space), which is UV divergent.
To deal with the divergence, it is suggested to take an effective
method. For smaller range of $r$ the Coulomb piece $1/r$ obviously
dominates, so that in $<\Psi|H_{Lamb}|\Psi>$ one can use the
wavefunction $\Psi$ which is the solution of the Schr\"odinger
equation containing only the Coulomb potential, namely we can have
an analytical solution for this asymptotic situation. Thus
$<\Psi|\ln {2\alpha\mu r}/r|\Psi>\propto \ln(na\mu/2)$. To make
the UV divergence vanish, the suggested renormalization scheme is
to set $\mu=2/\mathrm n a_B$. Indeed, in \cite{Titard:1993nn},
other three alternative schemes were also suggested, here we just
take this one and find the value of $\alpha_s$ determined with
this scheme is closer to that adopted in early literature for
calculating the spectra of bottomonia.

The value of the parameter $\Lambda$ is chosen as 0.2 GeV for
bottomonia\cite{Pineda:1998}, and at this point,
\begin{eqnarray}
 \alpha_s^{\mathrm n=2}=0.284
\end{eqnarray}
which is the value of $\alpha_s$ we used in Sec.\ \ref{sec-3}. It
is noted that $\alpha_s$ is different for different quantum number
$\mathrm n$:
\begin{eqnarray}\label{as}
 \alpha_s^{\mathrm n=1}=0.24~,~\alpha_s^{\mathrm n=2}=0.284~,
  ~\alpha_s^{\mathrm n=3}=0.316.
\end{eqnarray}
We will use the $\mathrm n$-related $\alpha_s$ value for
evaluating the spectra of the radially excited states of
bottomonia.

Simply adding the Lamb shift to the total binding energy is like
that we change the zero-point energy for each state. We still
select masses of $\Upsilon(1\mathrm S)$, $\chi_{b0}(1\mathrm P)$,
$\chi_{b1}(1\mathrm P)$, $\chi_{b2}(1\mathrm P)$,
$\Upsilon(2\mathrm S)$ as inputs, and solve the equation
(\ref{e-g}) again as we did in last section. But the value of
$\alpha_s$ in (\ref{e-g}) is taken as that given in Eq.(\ref{as})
which depends on  $\mathrm n$. The new solutions of $a$, $b$, $c$
and $d$ are:
\begin{eqnarray}\label{abcd-1}
 & &a^{(\mathrm{LM})}=1.4256,~b^{(\mathrm{LM})}=1.3553,\nonumber\\
 & &c^{(\mathrm{LM})}=0.8077,~d^{(\mathrm{LM})}=0.6849,
\end{eqnarray}
where the superscript LM refers to that all the corresponding
parameters are obtained as the Lamb shift being taken into
account.

With these solutions, our predictions on the whole family spectra
of bottomonia are presented in Tab.\ \ref{tab2}.

\begin{center}
 \begin{table}[!h]
 \caption{The mass spectra with the Lamb Shift (in GeV), where, the
 LM stands for the contribution of the Lamb Shift, $M$ is the
 predicted mass when the parameters are set as in Eq.(\ref{abcd-1})
 and $M'$ stands for $M'=M+\Delta E_\mathrm{LM}$.} \label{tab2}
 \def\temptablewidth{0.4\textwidth}
 {\rule{\temptablewidth}{0.5pt}}
  \begin{tabular*}{\temptablewidth}{@{\extracolsep{\fill}}lcccc}
   \toprule[1pt]
    meson &
      $\Delta E_\mathrm{LM}$ & $M$ &
       $M'$ & $M_\mathrm{EXP}$
    \\ \midrule[0.5pt]
    $\eta_b(1^1\mathrm S_0)$ &
     -0.1064 & 9.5274 & 9.4210 & 9.3020
    \\
    $\Upsilon(1^3\mathrm S_1)^\mathrm{fit}$ &
     -0.1114 & 9.5717 & 9.4603 & 9.4603
    \\
    $\chi_{b0}(1^3\mathrm P_0)^\mathrm{fit}$ &
     -0.0618 & 9.9212 & 9.8594 & 9.8594
    \\
    $\chi_{b1}(1^3\mathrm P_1)^\mathrm{fit}$ &
     -0.0620 & 9.9548 & 9.8928 & 9.8928
    \\
    $h_b(1^1\mathrm P_1)$ &
     -0.0621 & 9.9507 & 9.8887 &
    \\
    $\chi_{b2}(1^3\mathrm P_2)^\mathrm{fit}$ &
     -0.0622 & 9.9744 & 9.9122 & 9.9122
    \\
    $\eta_b(2^1\mathrm S_0)$ &
     -0.0549 & 10.0451 & 9.9902 &
    \\
    $\Upsilon(2^3\mathrm S_1)^\mathrm{fit}$ &
     -0.0561 & 10.0794 & 10.0233 & 10.0233
    \\
    $\Upsilon(1^3\mathrm D_1)$ &
     -0.0412 & 10.2139 & 10.1727 & 10.1611
    \\
    $\Upsilon(1^3\mathrm D_2)$ &
     -0.0412 & 10.2272 & 10.1860 &
    \\
    $\Upsilon(1^3\mathrm D_3)$ &
     -0.0412 & 10.2399 & 10.1987 &
    \\
    $\Upsilon(3^3\mathrm S_1)$ &
     -0.0379 & 10.4380 & 10.4001 & 10.3552
    \\
   \bottomrule[1pt]
  \end{tabular*}
  {\rule{\temptablewidth}{0.5pt}}
 \end{table}
\end{center}

\section{The spectra of the $b\bar c(\bar b c)$ mesons}\label{sec-5}

In this section, we further study the spectra of the $b\bar c(\bar
b c)$ mesons. Except the ground state $B_c$, the other states of
the $b\bar c$ mesons have not been well measured yet
\cite{Amsler:2008zzb}, we cannot directly fit the parameters from
data as what we do for the charmonia and the bottomonia. It is
noted that the parameters for charmonia and bottomonia are not
drastically apart and since the $b\bar c(\bar b c)$ family lies
between charmonia and the bottomonia, we may interpolate those
parameters for the $b\bar c(\bar b c)$ family, namely average the
values for charmonia and botomonia  to be that for the $b\bar
c(\bar b c)$ mesons (See the Tab.\ \ref{tab4}).

\begin{center}
 \begin{table}[!h]
  \caption{The parameters for $b\bar{c}$ (or $\bar{b}c$) mesons} \label{tab4}
  \def\temptablewidth{0.4\textwidth}
  {\rule{\temptablewidth}{0.5pt}}
   \begin{tabular*}{\temptablewidth}{@{\extracolsep{\fill}}cccccc}
    \toprule[1pt]
     meson &
       $a$ & $b$ &
        $c$ & $d$ & $\alpha_s$
    \\ \midrule[0.5pt]
    $c\bar{c}$\footnote{the parameters of the charmonia can be found in
\cite{Yuan:2010jy}} &
     1.1715 & 1.2250 & 0.8087 & 0.5291 & 0.36
    \\
    $b\bar{b}$\footnote{the parameters of the bottomonia are given in (\ref{abcd})} &
     1.2165 & 1.2988 & 0.8686 & 0.5886 & 0.284
    \\ \midrule[0.5pt]
    $b\bar{c}$ ( or $\bar{b}c$ ) &
     1.1940 & 1.2619 & 0.8387 & 0.5589 & 0.322
    \\
   \bottomrule[1pt]
  \end{tabular*}
  {\rule{\temptablewidth}{0.5pt}}
 \end{table}
\end{center}

Since the values of the $a\alpha_s$ and $c\alpha_s$ are more
useful for the calculation as discussed before, we re-define:
\begin{eqnarray}
 A=a\alpha_s\hspace{0.1in}\mbox{and}\hspace{0.1in}
  C=c\alpha_s.
\end{eqnarray}
Thus the parameters for the $b\bar c(\bar b c)$ mesons are:
\begin{eqnarray}\label{abcd-bc-ls}
 \left\{
  \begin{array}{l}
   A_{bc}=(A_b+A_c)/2;\hspace{0.1in}
    b_{bc}=(b_b+b_c)/2\\
   C_{bc}=(C_b+C_c)/2;\hspace{0.1in}
   d_{bc}=(d_b+d_c)/2.
  \end{array}
 \right.
\end{eqnarray}

The difference of the quark masses ($m_c=1.8\mathrm{GeV}$ and
$m_b=4.8\mathrm{GeV}$) make the Hamiltonian (\ref{B-F Hamilton})
to possess a more complicated form\cite{Lucha:1991it}:
\begin{subequations}\label{bc Hamilton}
 \begin{eqnarray}
  H&=&H_0+H_1+...\\
  H_0&=&{p^2\over2\mu}+m_b+m_c+V(r)+S(r)\\
  H_1&=&H_\mathrm{sd}+H_\mathrm{si}
 \end{eqnarray}
\begin{widetext}
 \begin{eqnarray}
  H_\mathrm{sd}&=&H_\mathrm{ls}+H_\mathrm{ss}+H_\mathrm{t}\nonumber\\
  &=&{1\over2r}\left(V'(r)-S'(r)\right)\left({{\bf L}\cdot {\bf S}_b\over
  m_b^2}+{{\bf L}\cdot {\bf S}_c\over m_c^2}\right)+{V'(r)\over m_b m_c}{\bf L}\cdot
  {\bf S}+{2\over3m_bm_c}\nabla^2V(r){\bf S}_b\cdot {\bf S}_c\nonumber\\
  & &+{1\over
  m_bm_c}\left({V'(r)\over r}-V''(r)\right)\left({({\bf S}_b\cdot r)({\bf S}_c\cdot
  r)\over r^2}-{1\over3}{\bf S}_b\cdot {\bf S}_c\right)\\
  H_\mathrm{si}&=&{1\over8}\left({1\over m_b^2}+{1\over m_c^2}-{2\over m_b
  m_c}\right)\nabla^2V(r)+{1\over4m_bm_c}\left\{{2\over
  r}V'(r)\cdot{\bf L}^2+\left[p^2,V(r)-rV'(r)\right]\right.\nonumber\\
  & &+\left.2\left(V(r)-rV'(r)\right)p^2+{1\over2}\left({8\over
  r}V'(r)+V''(r)-rV'''(r)\right)\right\}
 \end{eqnarray}
\end{widetext}
So the Schr\"odinger equation we need is:
 \begin{eqnarray}
  H\Psi(r)=(H_0+H_1)\Psi(r)=(E+m_b+m_c)\Psi(r)
 \end{eqnarray}
\end{subequations}
where,
\begin{eqnarray}
 \mu={m_bm_c\over m_b+m_c}
\end{eqnarray}

With this equation and the concerned parameters, we can predict
the spectra of the members of the whole $b\bar c(\bar b c)$ family
shown in Tab.\ \ref{tab5}.

\begin{center}
 \begin{table}[!h]
 \caption{The predict of the spectra of the $B_c$ meson} \label{tab5}
 \def\temptablewidth{0.5\textwidth}
 {\rule{\temptablewidth}{0.5pt}}
  \begin{tabular*}{\temptablewidth}{@{\extracolsep{\fill}}lcccc}
   \toprule[1pt]
    Quantity &
      our predict\footnote{with the parameter in Tab.\ \ref{tab4}}
       & our predict\footnote{Considering the effect of the Lamb shift
        and the parameters are taken as (\ref{abcd-bc-ls})}
         & EFG\cite{Godfrey:2004ya,Ebert:2002pp} &
         KWLC\cite{Ke:2010vx}
    \\ \midrule[0.5pt]
    $1^3S_1-1^1S_0$ & 0.0526 & 0.0448 &
     0.0620 & 0.0548
    \\
    $2^1S_0-1^1S_0$ & 0.5923 & 0.5843 &
     0.5650 & 0.5863
    \\
    $2^3S_1-1^3S_1$ & 0.5723 & 0.5744 &
     0.5430 & 0.5795
    \\
    $3^3S_1-2^3S_1$ & 0.4021 & 0.4067 &
     0.3540 & 0.3652
    \\
    $1^3P_0-1^3S_1$ & 0.3709 & 0.3885 &
     0.3670 &
    \\
    $1^1P_1-1^3P_0$ & 0.0492 & 0.0392 &
     0.0350 &
    \\
    $1^3P_1-1^3P_0$ & 0.0495 & 0.0411 &
     &
    \\
    $1^3P_2-1^1P_1$ & 0.0290 & 0.0320 &
     0.0280 &
    \\
    $1^3D_1-2^3S_1$ & 0.1341 & 0.1509 &
     0.1910 &
    \\
    $1^3D_2-1^3D_1$ & 0.0176 & 0.0253 &
     0.0050 &
    \\
    $1^3D_3-1^3D_2$ & 0.0170 & 0.0261 &
     0.0040 &
    \\
   \bottomrule[1pt]
  \end{tabular*}
  {\rule{\temptablewidth}{0.5pt}}
 \end{table}
\end{center}

\section{Conclusion and discussion}

In this work, we study the role of scalar potential to the spectra
of charmonia, bottomonia and the $b\bar c(\bar b c)$ family. Our
strategy is that the scalar and vector potentials have different
fractions which manifest in their coefficients (in the text, they
are $a$, $b$, $c$ and $d$ for the Coulomb and confinement pieces
respectively). By fitting some members of charmonia and bottomonia
which are more accurately measured, we fix them. Since except
$B_c$, the ground state of the $b\bar c(\bar b c)$ family, other
states have not been well measured yet, then we interpolate the
parameters for charmonia and bottomonia to determine the concerned
ones for the $b\bar c(\bar b c)$ mesons. With those parameters, we
further predict the mass spectra of the rest resonances of
charmonia, bottomonia and the whole $b\bar c(\bar b c)$ family. It
is shown that unlike the QED case where the fraction of the scalar
potential is very small and negligible, for the quarkonia where
QCD dominates, the fraction of scalar potential is of the same
order of magnitude as the vector potential. This is consistent
with the conclusion of Ref.\cite{Franklin:1998kd} and is not
surprising. As we indicated that for the vector-like coupling
theories QED and QCD, the scalar potential can only appear at loop
level or is induced by non-perturbative effect (QCD only). Thus it
should be loop-suppressed. However, for QCD, the coupling is
sizable and the higher order contributions and the
non-perturbative effects somehow are significant, so that one can
expect the fraction of the scalar potential is large.

Moreover, the Lamb shift is induced by the vacuum fluctuation and
only appears at loop level, indeed its leading contribution is at
$O(\alpha_s^2)$.  Therefore for the QED case, it is hard to
observe the Lamb shift (observation of the Lamb shift is a great
success for theory and experiment indeed), however, for QCD the
effects are not ignorable. It is shown
\cite{Zhang:2006ay,Gong:2009kp} that the NLO QCD effects may
exceed the LO contributions at some processes. By taking into
account of the Lamb shift, we re-fit the model parameters and find
they are obviously distinct from that without considering the Lamb
shift.

The results help us to get better understanding of QCD, higher
order effects and especially the non-perturbative effects. Even
though it is only half-quantitative, it is an insight to the whole
picture.

In this work, we adopt the renormalization scheme as
$\mu={2/\mathrm na_B}$ \cite{Titard:1993nn}, which determines the
effective coupling $\alpha_s$. It is worth emphasizing that
$\alpha_s$ depends on the principal quantum number $\mathrm n$ and
this is different from that usually used in literature. But for
the ground state of charmonia and bottomonia, the values of
$\alpha_s$ are quite close to that appearing in the literature.

The predictions on the $b\bar c(\bar b c)$ family will be tested
at LHCb experiments where a great amount of the excite states of
$b\bar c(\bar b c)$ will be produced. Comparing with the data, we
will learn more about the QCD and structures of the ``final''
meson family.

Actually, in this work, we only consider the Cornell potential
which is supported by the area theorem and commonly adopted in
phenomenological studies where the spectra and wavefunctions of
heavy mesons are involved. Indeed, there are some other proposals.
For example, the authors of Ref. \cite{Beveren} use the harmonic
oscillator model to deal with the confinement and further consider
the effects of open charm loop for higher excited charmonia states
which may induce energy shifts and change decay widths. Moreover,
some authors introduce a phenomenological form for the spin-spin
interaction \cite{Eichten} which may also result in energy level
shift. In this work, we just restrict ourselves at the quark level
QCD motivated potential whose form is given in
Ref.\cite{Lucha:1991it,Lucha:1991vn}, but we may further our
studies on the coupled-channel scenario in our coming work.

The contribution of the scalar potential to the hadron spectra was
noticed by some authors \cite{Barchielli:1988zp} long time ago,
and its importance was confirmed. In our present work, we
re-emphasize its role and discuss the origin in comparison with
the QED case. In terms of the newly achieved data on charmonia and
bottomonia, we analyze the hadron spectra and gain all the
concerned parameters. We also investigate the significance of the
Lamb shift phenomenologically. Then we go on discussing the
spectra of $b\bar c (\bar bc)$ family within the same framework,
the results will be tested in the future experiments

\section*{Acknowledgments}
This project is supported by the National Natural Science
Foundation of China (NSFC) under Contracts No. 10775073 and No.
11005079; the Special Grant for the Ph.D. program of Ministry of
Eduction of P.R. China No. 20070055037 and No. 20100032120065.

\vspace {2cm}

{\bf Appendix : Check the legitimacy of the approximation
$p^4\sim\Big[2\mu\left(E_0-V(r)\right)\Big]^2$}\\

We investigate and elucidate the legitimacy of the approximation
adopted  in the text:
$$p^4\sim
 \Big[2\mu(E_0-V(r))\Big]^2$$
through a few examples.

In fact, such problems have been thoroughly discussed in
literature and even written in textbooks, for example in
Ref.\cite{Claude} and the relativistic corrections to the Cornell
potential can be found in Ref.\cite{McClary:1983xw}. Here we just
re-do the numerical computation to convince our readers and
ourselves about the legitimacy of the approximation adopted in the
text because it is very important for getting the spectra.

The 0-th order Schr\"ordinger equation is:
 \begin{eqnarray}\label{appendix-1}
  \left[{p^2\over2\mu}+V(r)\right]\Psi=E_0\Psi
 \end{eqnarray}
 With the relativistic correction, it will be:
 \begin{eqnarray}\label{appendix-2}
  \left[{p^2\over2\mu}+V(r)-{p^4\over4m^3}\right]\Psi=E_1\Psi
 \end{eqnarray}
 Note: here we ignore other irrelevant correction terms such as the
 L-S coupling etc. because we only concern the $p^4$ term and the
 approximation.

 First, let us use the Coulomb potential as an example because there
 exists an analytical solution.

 For the Coulomb potential:
 \begin{eqnarray}
  V(r)=-{a\over r}
 \end{eqnarray}
 We have the exact solution (the eigen-energy and the wave function):
 \begin{subequations}
  \begin{eqnarray}
   E_0^\mathrm{n}&=&-{a^2\over4\mathrm n^2}\\
   R_{10}(r)&=&2K^{3\over2}e^{-Kr}\\
   R_{20}(r)&=&\left({1\over2}K\right)^{3\over2}\left(2-Kr\right)e^{-{1\over2}Kr}\\
   R_{30}(r)&=&\left({1\over3}K\right)^{3\over2}\left[2-{4\over3}K
    +{4\over27}K^2\right]e^{-{1\over3}Kr}
  \end{eqnarray}
  where,
  \begin{eqnarray}
    \mu&=&{m^2\over2m}={m\over2};\hspace{0.3in}
     K={1\over2}ma\nonumber
  \end{eqnarray}
 \end{subequations}
If $m=1.84$ GeV, and $a=0.5$ (these numbers are just taken for an
illustration, but not for real physics), then:
 \begin{eqnarray}
  \left\{
   \begin{array}{l}
    E_0^{\mathrm n=1}=-0.115\\
    E_0^{\mathrm n=2}=-0.02875\\
    E_0^{\mathrm n=3}=-0.012778
   \end{array}
  \right.
 \end{eqnarray}
 and in the perturbative method, the relativistic correction is:
 \begin{eqnarray}\label{appendix-a1}
  \Delta E^\mathrm{n}&=&\int
   dr~r^2{1\over4m^3}\left[{\bf p}^2R_{\mathrm n0}(r)\right]^2\nonumber\\
  \Rightarrow & &
  \left\{
   \begin{array}{l}
    \Delta E^{\mathrm n=1}=0.008984;\\
    \Delta E^{\mathrm n=2}=0.001460;\\
    \Delta E^{\mathrm n=3}=0.000558.
   \end{array}
  \right.
 \end{eqnarray}

On the other hand, as we use the approximation:
$p^4\sim\Big[2\mu\left(E_0-V(r)\right)\Big]^2$ in
Eq.(\ref{appendix-2}), \footnote{here, we may use $E_0$ as well,
but for a clear comparison we use $E_1$ instead. The error is not
great.} and then find the numerical solution of Eq.
(\ref{appendix-2}):
 \begin{eqnarray}
  \left\{
   \begin{array}{l}
    E_1^{\mathrm n=1}=-0.1223;\\
    E_1^{\mathrm n=2}=-0.0300;\\
    E_1^{\mathrm n=3}=-0.0130.
   \end{array}
  \right.
 \end{eqnarray}
 If we define: $\Delta E_1=E_0^{\mathrm n}-E_1^{\mathrm n}$, then we have:
 \begin{eqnarray}\label{appendix-a2}
  \left\{
   \begin{array}{l}
    \Delta E_1^{\mathrm n=1}=0.007265;\\
    \Delta E_1^{\mathrm n=2}=0.001249;\\
    \Delta E_1^{\mathrm n=3}=0.0002288.
   \end{array}
  \right.
 \end{eqnarray}

From the Eqs.(\ref{appendix-a1}) and (\ref{appendix-a2}), we can
find that the error is  near 0.001 GeV. The relative error is:
 \begin{eqnarray}
  x^\mathrm{n}&=&{\left(E_0^\mathrm{n}+\Delta E^\mathrm{n}\right)
   -E_1^\mathrm{n}\over \left(E_0^\mathrm{n}+\Delta E^\mathrm{n}\right)+E_1^\mathrm{n}}
    \nonumber\\
  \Rightarrow & &
  \left\{
   \begin{array}{l}
    x^{\mathrm n=1}=0.68\%;\\
    x^{\mathrm n=2}=0.35\%;\\
    x^{\mathrm n=3}=1.28\%.
   \end{array}
  \right.
 \end{eqnarray}

With the Coulomb potential the Schr\"odinger equation possesses an
analytical solution, so it is easy to see the error. However, for
the Cornell potential,  there is no analytical solution available,
so that we need to use the numerical solution for our analysis.

The Cornell potential is:
 \begin{eqnarray}
  V(r)=-{a\over r}+br
 \end{eqnarray}
where, we set $a=0.5$, $b=0.2$  and consider the second radially
excited state with $\mathrm n=2$ for an example. The numerical
solution of Eq.(\ref{appendix-1}) is:
 \begin{eqnarray}
  E_0^{\mathrm n=2}=0.856872
 \end{eqnarray}
 and the numerical solution of Eq.(\ref{appendix-2}) is:
 \begin{eqnarray}
  E_1^{\mathrm n=2}=0.918242
 \end{eqnarray}
 So we have:
 \begin{eqnarray}\label{appendix-a3}
  \Delta E^{\prime\mathrm n=2}=E_1^{\mathrm n=2}-E_0^{\mathrm n=2}=0.06137
 \end{eqnarray}
In the numerical solution of Eq.(\ref{appendix-1}) we have the
wave function $R_\mathrm {nl}(r)$, and with the perturbative
method, the contribution of the relativistic correction is:
 \begin{eqnarray}\label{appendix-a4}
  \Delta E^{\mathrm n=2}&=&
   \int {1\over4m^3}\left[{\bf p}^2R_{20}(r)\right]^2 r^2dr\nonumber\\
  &=&0.0669..
 \end{eqnarray}
Finally, from Eqs.(\ref{appendix-a3}) and (\ref{appendix-a4}), we
have the relative error:
 \begin{eqnarray}
  x^{\mathrm n=2}&=&{\left(E_0^{\mathrm n=2}+\Delta E^{\mathrm n=2}\right)
   -E_1^{\mathrm n=2}\over \left(E_0^{\mathrm n=2}+\Delta
  E^{\mathrm n=2}\right)-E_1^{\mathrm n=2}}=0.30\%.
 \end{eqnarray}
Even though the relativistic error seems large, in fact the error
is only at order of a few of MeV. The QCD Lamb shift generally
results in a few of tens of MeV, thus the error brought up by the
approximation seems not too serious.

Here one important point should be clarified. Directly calculating
Eq.(\ref{appendix-a4}), one would have an unrealistically large
result. The reason is obvious that unlike the analytical solution
for the Coulomb potential, the behavior of the numerical solution
near the zero point ($r\to 0$) is not appropriate (namely it does
not possess an exponential factor to guarantee the convergence of
the solutions). In other words, as $r\to 0$ corresponding
$p\to\infty$, the contribution of the power term $p^m\; (m>2)$
would become larger and larger for higher $m$. Indeed all higher
power terms should exist in the relativistic corrections.
Integration over the wavefunction would blow up. Thus the whole
picture is not acceptable. To remedy it, there are two ways, one
is introducing an exponential convergence factor and another is
restricting the integration region, i.e. one does not integrate
from $r=0$, but set the lower bound to be a small number $\delta$.
Definitely, one should find a small value of $\delta$ and when its
value changes slightly, the result does not vary much. Thus we
would convince ourselves that this integration is reliable. The
second way is much simpler than the first one, and we adopt it for
the above calculation. We used to make a careful discussions on
such virtual singularity in our earlier paper\cite{Ding:1999hy}.


\vspace{1cm}

\begin{thebibliography}{}

\bibitem{Quigg:1979vr}
  C.~Quigg and J.~L.~Rosner,
  Phys.\ Rept.\  {\bf 56}, 167 (1979).

\bibitem{Brambilla:2004wf}
  N.~Brambilla {\it et al.}  [Quarkonium Working Group],
  arXiv:hep-ph/0412158.

\bibitem{chen:2008}
  F.~Zhang,\ B.~Fu and J.~Chen,
  Phys.\ Rev.\  A {\bf 78}, 040101(R) (2008).

\bibitem{Ke:2009mx}
  H.~W.~Ke, Z.~Li, J.~L.~Chen, Y.~B.~Ding and X.~Q.~Li,
  Int.\ J.\ Mod.\ Phys.\  A {\bf 25}, 1123 (2010)
  [arXiv:0907.0051 [hep-ph]].

\bibitem{Zhang:2006ay}
  Y.~J.~Zhang and K.~T.~Chao,
  Phys.\ Rev.\ Lett.\  {\bf 98}, 092003 (2007)
  [arXiv:hep-ph/0611086].

\bibitem{Gong:2009kp}
  B.~Gong and J.~X.~Wang,
  Phys.\ Rev.\ Lett.\  {\bf 102}, 162003 (2009)
  [arXiv:0901.0117 [hep-ph]].


\bibitem{Leviatan:2003wy}
  A.~Leviatan,
  Phys.\ Rev.\ Lett.\  {\bf 92}, 202501 (2004)
  [Erratum-ibid.\  {\bf 92}, 219902 (2004)]
  [arXiv:nucl-th/0312018];\
  Int.\ J.\ Mod.\ Phys.\  E {\bf 14}, 111 (2005)
  [arXiv:nucl-th/0407107];\
  Phys.\ Rev.\ Lett.\  {\bf 103}, 042502 (2009)
  [arXiv:0907.3557 [nucl-th]].

\bibitem{Yuan:2010jy}
  X.~H.~Yuan, H.~W.~Ke and X.~Q.~Li,
  arXiv:1011.4629 [hep-ph].

\bibitem{Lucha:1991it}
  W.~Lucha, H.~Rupprecht and F.~F.~Schoberl,
  Phys.\ Rev.\  D {\bf 46}, 1088 (1992).

\bibitem{Lucha:1991vn}
  W.~Lucha, F.~F.~Schoberl and D.~Gromes,
  Phys.\ Rept.\  {\bf 200}, 127 (1991).

\bibitem{Ding:1991vu}
  Y.~B.~Ding, D.~H.~Qin and K.~T.~Chao,
  Phys.\ Rev.\  D {\bf 44}, 3562 (1991).

\bibitem{Hoang:2001rr}
  A.~H.~Hoang, A.~V.~Manohar and I.~W.~Stewart,
  Phys.\ Rev.\  D {\bf 64}, 014033 (2001)
  [arXiv:hep-ph/0102257].

\bibitem{Titard:1993nn}
  S.~Titard and F.~J.~Yndurain,
  Phys.\ Rev.\  D {\bf 49}, 6007 (1994)
  [arXiv:hep-ph/9310236];\
    Phys.\ Rev.\ D.\ {\bf 51},\ 6348 (1995)


\bibitem{Brambilla:1999xf}
  N.~Brambilla, A.~Pineda, J.~Soto and A.~Vairo,
  Nucl.\ Phys.\  B {\bf 566}, 275 (2000)
  [arXiv:hep-ph/9907240];
  N.~Brambilla, A.~Pineda, J.~Soto and A.~Vairo,
  Rev.\ Mod.\ Phys.\  {\bf 77}, 1423 (2005)
  [arXiv:hep-ph/0410047].





\bibitem{Ke:2009sk}
  H.~W.~Ke and X.~Q.~Li,
  Sci.\ China {\bf G53}, 2019 (2010)
  [arXiv:0910.1158 [hep-ph]].

\bibitem{Eichten:cornell}
  E.~Eichten {\it et al.},
   Phys.\ Rev.\ Lett.\ {\bf 34}, 369 (1975);
   Phys.\ Rev.\ D.\ {\bf 17}, 3090 (1978);
   Phys.\ Rev.\ D.\ {\bf 21}, 203 (1980);

\bibitem{Cai:2003}
  C.\ H.\ Cai,~and~L.\ Lei,
   HEP \& NP {\bf 27}(11), 1005 (2003).

\bibitem{Silbar:2010}
  R.\ Silbar,~and~T.\ Goldman,
   [arXiv:1001.2514v1].

\bibitem{Amsler:2008zzb}
  C.~Amsler {\it et al.}  [Particle Data Group],
  Phys.\ Lett.\  B {\bf 667}, 1 (2008).

\bibitem{Press:2007}
  William\ H.\ Press {\it et al. },\
    {\it Numerical recipes: the art of scientific computing},\
    Cambridge University Press\ (2007).

\bibitem{Ding:1988pg}
  H.~Q.~Ding,
  Phys.\ Lett.\  B {\bf 200} (1988) 133.

\bibitem{Bali:1992}
  G.\ S.\ Bali~and~K.\ Schilling,
   Phys.\ Rev.\ D.\ {\bf 46}, 2636 (1992).

\bibitem{Greiner:1992bv}
  W.\ Greiner and J.\ Reinhardt,
{\it Quantum Electrodynamics}, Berlin, Germany: Springer (1992).

\bibitem{Pineda:1998}
  A.\ Pineda ,\ and \ F.\ J.\ Yndurain,\
    Phys.\ Rev.\ D.\ {\bf 58},\ 094022 (1998);
    Phys.\ Rev.\ D.\ {\bf 61},\ 077505 (2000).

\bibitem{Billoire:1979ih}
  A.~Billoire,
  Phys.\ Lett.\  B {\bf 92}, 343 (1980).

\bibitem{Fischler:1977yf}
  W.~Fischler,
  Nucl.\ Phys.\  B {\bf 129}, 157 (1977).

\bibitem{Peter:1996ig}
  M.~Peter,
  Phys.\ Rev.\ Lett.\  {\bf 78}, 602 (1997)
  [arXiv:hep-ph/9610209].

\bibitem{Schroder:1998vy}
  Y.~Schroder,
  Phys.\ Lett.\  B {\bf 447}, 321 (1999)
  [arXiv:hep-ph/9812205];\

  H.~W.~Crater, J.~H.~Yoon and C.~Y.~Wong,
  Phys.\ Rev.\  D {\bf 79}, 034011 (2009).

\bibitem{Godfrey:2004ya}
  S.~Godfrey,
  Phys.\ Rev.\  D {\bf 70}, 054017 (2004)
  [arXiv:hep-ph/0406228].

\bibitem{Ebert:2002pp}
  D.~Ebert, R.~N.~Faustov and V.~O.~Galkin,
  Phys.\ Rev.\  D {\bf 67}, 014027 (2003)
  [arXiv:hep-ph/0210381].

\bibitem{Ke:2010vx}
  H.~W.~Ke, G.~L.~Wang, X.~Q.~Li and C.~H.~Chang,
  Sci.\ China {\bf G53}, 2025 (2010)
  [arXiv:1002.4051 [hep-ph]].

\bibitem{Franklin:1998kd}
  J.~Franklin,
  Mod.\ Phys.\ Lett.\  A {\bf 14}, 2409 (1999);

  A.~S.~de Castro and J.~Franklin,
  arXiv:hep-ph/0011137;
  Int.\ J.\ Mod.\ Phys.\  A {\bf 15}, 4355 (2000);

\bibitem{Beveren}
  E.~van Beveren, C.~Dullemond and G.~Rupp,
  Phys.\ Rev.\  D {\bf 21}, 772 (1980)
  [Erratum-ibid.\  D {\bf 22}, 787 (1980)];

  E.~van Beveren and G.~Rupp,
  arXiv:1009.1778 [hep-ph];
  arXiv:1009.3395 [hep-ph].




\bibitem{Eichten}
  E.~Eichten, K.~Gottfried, T.~Kinoshita, K.~D.~Lane and T.~M.~Yan,
  Phys.\ Rev.\  D {\bf 17}, 3090 (1978)
  [Erratum-ibid.\  D {\bf 21}, 313 (1980)];\
  Phys.\ Rev.\  D {\bf 21}, 203 (1980).

\bibitem{Claude}
 Claude Cohen-Tannoudji,\ Bernard Diu,\ Frank Laloe,\
  {\it Quantum Mechanics, 2 Volume Set}, John Wiley \& Sons, Inc. (2006).

\bibitem{McClary:1983xw}
  R.~McClary and N.~Byers,
  Phys.\ Rev.\  D {\bf 28}, 1692 (1983).

\bibitem{Ding:1999hy}
  Y.~B.~Ding, X.~Q.~Li and P.~N.~Shen,
  Commun.\ Theor.\ Phys.\  {\bf 33}, 613 (2000)
  [arXiv:hep-ph/9902330].

\bibitem{Barchielli:1988zp}
  A.~Barchielli, N.~Brambilla and G.~M.~Prosperi,
  Nuovo Cim.\  A {\bf 103}, 59 (1990).



\end{thebibliography}
\end{document}